\documentclass[11pt]{article}

\usepackage{amsmath, amssymb, latexsym}
\usepackage{textcomp}

\newtheorem{thm}{\bf Theorem}[section]
\newtheorem{prop}[thm]{\bf Proposition}  

\newtheorem{remark}[thm]{\bf Remark}

\newcommand{\Bbox}{
{\unskip\nobreak\hfil\penalty50
\hskip1em\hbox{}\nobreak\hfil{\lower .5pt \hbox{$\Box$}}
\parfillskip=0pt \finalhyphendemerits=0 \par}
}

\newcommand{\eop}{
\ifmmode {\hbox{\Bbox}} \else \Bbox \fi
}

\newcommand{\zos}{\{0,1\}^*}

\newcommand{\lex}{<_\ell}

\newcommand{\Q}{\mathbb{Q}}

\title{An undecidable property of context-free languages}
\author{
Z. \'Esik\thanks{Partially supported by grant no. K 75249
from the National Foundation of Hungary for Scientific Reseach (OTKA).} \\
Department of Informatics \\
University of Szeged\\
Szeged, Hungary}

\date{April 7, 2010}
\begin{document}
\maketitle

\begin{abstract}
We prove that there exists no algorithm to decide whether 
the language generated by a context-free grammar 
is dense with respect to the lexicographic ordering. 
As a corollary to this result, we show that it 
is undecidable whether the lexicographic orderings 
of the languages generated
by two context-free grammars have the same order type.
\end{abstract}

\section{Introduction}

Suppose that $\Sigma$ is an alphabet equipped with a (strict) 
linear order relation $<$. We may extend $<$ to a lexicographic 
ordering $<_\ell$ of $\Sigma^*$ by defining, for all words $u,v \in \Sigma^*$, 
$u <_\ell v$ if either $u$ is a proper prefix of $v$, or $u = xay$
and $v = xbz$ for some $a,b \in \Sigma$ and $x,y,z \in \Sigma^*$
with $a < b$. Thus, when $L \subseteq \Sigma^*$, then $(L,<_\ell)$ 
is a linear ordering. It is known (see e.g. \cite{BEbergen,Courcelle78}) that
if the size of $\Sigma$ is two or more, then every  
countable linear ordering is isomorphic to a linear 
ordering $(L,<_\ell)$ for some language $L \subseteq \Sigma^*$. 
Let us call a linear ordering \emph{regular, context-free, 
or deterministic context-free} if it is isomorphic to 
the linear ordering of a language of the appropriate type.

It follows by the characterization of regular and algebraic 
trees by their branch languages \cite{Courcelle78,Courcelle} 
that the regular (deterministic context-free) 
linear orderings are exactly those that can be defined by 
recursion schemes of order $0$ (order $1$, respectively).
See also  \cite{BEbergen}. Moreover, a well-ordering is regular
if and only if its order type is less than $\omega^\omega$, and 
deterministic context-free if and only if its order type is less than
$\omega^{\omega^\omega}$, cf. \cite{BEordinals}. (These well-orderings  
have other characterizations using operations on well-orderings or automata, 
cf. \cite{Delhomme,Khoussainovetal}.)
Moreover, it follows from results proved in 
\cite{Heilbrunner} that the Hausdorff rank \cite{Rosenstein}
of every scattered regular linear ordering is finite. 
As shown in \cite{BEscattered}, the Hausdorff rank of 
every scattered deterministic 
context-free linear ordering is less than $\omega^\omega$.
Ordinals and scattered linear orderings defined by 
higher order recursion schemes are studied in
\cite{BraudCarayol}.

It was shown in \cite{Thomas} that it is decidable for 
regular linear orderings (given as lexicographic orderings
of regular languages) whether they are isomorphic. 
The decidability status of the isomorphism problem for 
deterministic context-free linear orderings is open. Here, we show 
that it is undecidable for context-free linear orderings (given
by context-free grammars) whether they are isomorphic. Moreover, we show that 
it is undecidable whether a context-free 
language defines a dense linear ordering.

\section{Linear orderings and context-free grammars}

A \emph{linear ordering} \cite{Rosenstein} is a set $S$ equipped with a strict 
linear order relation $<$. In this paper, we restrict ourselves 
to linear orderings $(S,<)$, where $S$ is a countable set.
A linear ordering $(S,<)$ is \emph{dense} if it has at least 
two elements and for any $x,y \in S$ with $x < y$ 
there is some $z$ with $x < z < y$. Two 
linear orderings $(S,<)$ and $(S',<)$ are 
isomorphic if there is a bijection $h: S \to S'$ 
such that $xh < yh$ for all $x,y \in S$ with $x < y$.
Isomorphic linear orderings have the same \emph{order type}.
It is known that up to isomorphism there are 4 dense 
(countable) linear orderings, the ordering $\Q$ of the rationals
possibly equipped with a least or greatest element (or both). 
The order type of $\Q$ is denoted $\eta$.

A \emph{context-free grammar} $G$ over a (terminal) alphabet $\Sigma$ consists of 
a finite nonempty set $N$ of nonterminals and a finite 
set of productions $A \to u$, where $A \in N$ and 
$u \in (N \cup \Sigma)^*$. It is assumed that 
$N$ and $\Sigma$ are disjoint. A nonterminal $A_0$,
called the start symbol, is distinguished. The derivation relation $\Rightarrow^*$ 
is defined as usual. For each nonterminal $A$, we let $L(G,A)
= \{ u \in \Sigma^* : A \Rightarrow^* u\}$ denote the language 
generated from $A$. The \emph{context-free language} $L(G) \subseteq \Sigma^*$ 
generated by $G$ is $L(G,A_0)$. We call $G$ a \emph{prefix grammar}
if the languages $L(G,A)$ are all prefix (or prefix-free) languages.
A \emph{right linear grammar} is a context-free grammar such that,
except possibly for the last letter, each letter occurring in the word 
on the right side of a production is a terminal letter. It is 
well-known that a language is \emph{regular} if and only if it can be generated
by a right-linear grammar. For all unexplained notions on context-free 
grammars and languages refer to any standard book on formal languages. 

The reverse of a word $u$ will be denoted $u^{-1}$. 

\begin{remark}
It was pointed out by Luc Boasson that there is no algorithm to decide 
for a context-free grammar $G$ whether it is a prefix grammar.
Moreover, there is no algorithm to decide whether a given
context-free grammar
generates a prefix language.
\end{remark}

\section{Some undecidability results}

In this section our aim is to prove that it is undecidable for a 
context-free (prefix) grammar $G$ over a 2-letter alphabet 
whether or not $(L(G),\lex)$ is a dense ordering, or a linear 
ordering isomorphic to the ordering $\Q$ of the rationals. It follows 
from this result that it is undecidable whether or not the lexicographic 
orderings of two context-free languages, given by context-free (prefix) 
grammars, are isomorphic. In our proofs, we will use reduction from 
the Post Correspondence Problem (PCP). 

Let $(\alpha,\beta)$ be an instance of PCP, where  
$\alpha = (\alpha_1,\ldots,\alpha_n)$
and $\beta = (\beta_1,\ldots,\beta_n)$ 
are nonempty  sequences of nonempty words over the 
two-letter alphabet  $\{a,b\}$. 
Then consider the alphabet 
$$\Gamma = \{1,\ldots,n,a,b,\hbox{\textcent},\$ \},$$
ordered as indicated. For convenience, 
we will also refer to the elements of $\Gamma$ 
by the letters $c_1,c_2,\ldots,c_{n+4}$
with $c_1$ denoting $1$, $c_2$ denoting $2$, etc. 
For $j = 1,\ldots, n+2$, define 
$\Delta_j$ as the 3-letter alphabet $\{d_{j0},d_{j1},d_{j2}\}$
and extend the linear order on $\Gamma$ to a linear ordering of the set 
$$\Delta = \Gamma \cup \bigcup_{j = 1}^{n+2} \Delta_j$$
so that 
$$c_j < d_{j0} < d_{j1} < d_{j2} < c_{j+1}$$
for all $j = 1,\ldots,n+2$.
Note that $\Delta$ contains $4n + 10$ letters 
and there is no ``extra letter'' between $\hbox{\textcent}$ and $\$ $.

We will construct a (prefix) grammar $G = G_{\alpha,\beta}$
over the alphabet $\Delta$ such that $(L(G),\lex)$ 
is dense if and only if $(\alpha,\beta)$ has no solution. 
The grammar $G$ will be designed so that it will generate
the language
  $$L = L_\alpha \cup L_\beta \cup L_1 \cup \ldots \cup L_{n+2}$$
where
\begin{enumerate}
\item $L_\alpha = \{i_1\ldots i_m(\alpha_{i_1}\ldots\alpha_{i_m})^{-1}\hbox{\textcent} : 1 \leq i_k \leq n,\ m \geq 1\}$
\item $L_\beta = \{i_1\ldots i_m(\beta_{i_1}\ldots\beta_{i_m})^{-1}\$ : 1 \leq i_k \leq n,\ m \geq 1\}$
\item $L_j = \{1,\ldots,n,a,b\}^*Q_j$, where $Q_j = \{d_{j0},d_{j2}\}^* d_{j1}$, $j = 1,\ldots,n+2$.
\end{enumerate} 
Note that each $Q_j$ and each $L_j$ is a dense regular language whose order type 
is $\eta$, the order type of the rationals. The same fact holds for the languages 
$Q = \bigcup_{j =1}^{n+2} Q_j$ and $L' = \bigcup_{j =1}^{n+2} L_j$, since 
the order type of any finite nonempty sum $\sum_{i \in I}P_i$ 
of linear orderings $P_i$ of order type $\eta$ is also $\eta$. 

The grammar $G$ has start symbol $S$ and contains the following productions in BNF:
\begin{eqnarray*}
S & \to & A\hbox{\textcent} \mid B\$ \mid C\\
A & \to & iA\alpha_i^{-1} \mid i\alpha_i^{-1}\\
B & \to & iB\beta_i^{-1} \mid i\beta_i^{-1}\\
C &\to & iC \mid aC \mid bC \\
C & \to & D_1 \mid \ldots \mid D_{n+2}\\
D_j &\to & d_{j0}D_j \mid d_{j2}D_j \mid d_{j1}
\end{eqnarray*}
It is clear that $G$ is a prefix grammar.

\begin{prop}
\label{prop-undec1}
$(L(G_{\alpha,\beta}),\lex)$ is dense if and only if $(\alpha,\beta)$ has no solution. 
\end{prop}

{\sl Proof.} Assume that $i_1\ldots i_m$ is a solution of $(\alpha,\beta)$.
 Let $u =(\alpha_{i_1}\ldots \alpha_{i_m})^{-1}
= (\beta_{i_1}\ldots \beta_{i_m})^{-1}$. Then 
$$u_\alpha = i_1\ldots i_m u\hbox{\textcent}\quad {\rm and}\quad u_\beta= i_1\ldots i_m u\$$$
are in $L$. However, there is no word $v$ in $L$  with 
$$u_\alpha \lex v \lex u_\beta,$$ showing that $L$ is not dense. 

Suppose now that $(\alpha,\beta)$ has no solution. We show that 
$L$ is dense. To this end, suppose that $u,v\in L$ with $u \lex v$.
Since $L$ is a prefix language, $u$ and $v$ can be decomposed as
$$u = wcu',\ v = wdv'$$ where $c$ and $d$ are letters with $c < d$.
It is not possible that $c = \hbox{\textcent}$ and $d = \$ $, since otherwise we would have 
$u' = v' = \epsilon$ and the maximal prefix of $w$ that is in $\{1,\ldots,n\}^*$
would give a solution of $(\alpha,\beta)$.

Thus, either $c \in \Delta_i$ or $c = c_i$ for some $i = 1,\ldots,n+2$. 
There are three cases to consider.  
\begin{enumerate}
\item $c \in \Delta_i$ for some $i = 1,\ldots,n+2$, so that 
$cu' \in Q_i$. If $d$ is also in $\Delta_i$, then 
$dv' \in Q_i$, and since $cu' \lex dv'$, there exists
some $x \in Q_i$ with $cu' \lex x \lex dv'$ and thus
$u = wcu' \lex wx \lex wdv'$, where $wx$ is in $L$. If $d \not\in \Delta_i$ 
then choose any word $x \in Q_i$ with $cu' \lex x$.
We again have $u = wcu' \lex wx \lex wdv'$ and $wx \in L$. 
\item $d \in \Delta_i$ for some $i = 1,\ldots,n+2$. This case is symmetrical to 
the previous case. 
\item Thus the only remaining case is when $c = c_i$ for some $i = 1,\ldots,n+2$
and $d = c_j$ for some $j = 1,\ldots,n+4$ with $i < j$. 
In this case let $x$ be any word in $Q_i$. We have that $u = wcu' \lex wx \lex wdv'$
and $wx \in L$.
\end{enumerate} 
Thus, we have shown that if $(\alpha,\beta)$ has no solution,
then between any two words of $L$ there is a third word of $L$,
completing the proof of the fact that $L$ is dense. \eop 

\begin{remark}
\label{rem-undec1}
The language $L= L(G_{\alpha,\beta})$ generated by the above grammar $G_{\alpha,\beta}$ 
has no least or greatest element with respect to the lexicographic order. Indeed,
if $v \in L'$, then there exist words $u,w \in L'$ with $u \lex v \lex w$ since
the order type of $L'$ is $\eta$. Now consider a word 
$v = i_1\ldots i_m(\alpha_{i_1}\ldots\alpha_{i_m})^{-1}\hbox{\textcent}$ in $L_\alpha$. 
Then let $u = i_1\ldots i_m1 (\alpha_{i_1}\ldots\alpha_{i_m}\alpha_1)^{-1}\hbox{\textcent}$ 
and let $w = d_{i_11}$ or any other word in $Q_{i_1}$.  We have that $u \lex v \lex w$ and $u,w \in L$.  
Similarly, if  $v = i_1\ldots i_m (\beta_{i_1}\ldots\beta_{i_m})^{-1}\$$ is in $L_\beta$
then $u \lex v \lex w$ for the words
$u = i_1\ldots i_m1 (\beta_{i_1}\ldots\beta_{i_m}\beta_1)^{-1}\$$
and $w = d_{i_11}$ in $L$.
\end{remark} 

We order the binary alphabet $\{0,1\}$ by $0 < 1$.

\begin{thm}
There exists no algorithm to decide for a context-free (prefix) grammar $G$ over $\{0,1\}$ 
whether $(L(G),\lex)$ is dense. Moreover, there exists no algorithm to decide
for  a context-free (prefix) grammar $G$ over $\{0,1\}$ whether the order type of $(L(G), \lex)$
is $\eta$.  
\end{thm}

{\sl Proof.} This follows from Proposition~\ref{prop-undec1} and Remark~\ref{rem-undec1}
by an appropriate order preserving coding of the letters of the alphabet $\Delta$
by words over $\zos$ of length $\lceil \log (4n + 10)\rceil$.
\eop

\begin{thm}
There exists no algorithm to decide for a context-free (prefix) grammar $G$ 
and a right linear (prefix) grammar $G'$ over $\{0,1\}$  whether $(L(G),\lex)$ and 
$(L(G'), \lex)$ are isomorphic. 
\end{thm}

{\sl Proof.} Consider an instance $(\alpha,\beta)$ of PCP and the grammar $G = G_{\alpha,\beta}$ 
constructed above. As before, let us code terminal letters by words 
of length  $\lceil \log (4n + 10)\rceil$ by an order preserving coding. 
Thus, $L(G)$ is a language over the alphabet $\zos$ such that the order type 
of $(L(G),\lex)$ is $\eta$ if and only if $(\alpha,\beta)$ has no solution. Then
let $G'$ be the right linear (prefix) grammar with productions 
$$S \quad \to \quad 00S \mid 11S \mid 01$$
generating the language $\{00,11\}^*01$ of order type $\eta$. 
Then $(L(G),\lex)$ and $(L(G'),\lex)$ are isomorphic if and only if 
$(\alpha,\beta)$ has no solution.
\eop 

\section{Conclusion}

We have proved that there is no algorithm to decide whether 
a context-free grammar (even prefix grammar) generates a dense 
language with respect to the lexicographic ordering. 
As a corollary to this result, we have shown that it 
is undecidable whether two prefix grammars generate languages 
of the same order type.

We can prove that it is decidable in polynomial 
time whether the lexicographic ordering of the language 
generated by a prefix grammar is scattered, or a well-ordering.
Moreover, we can extend the decidability part of this result
to arbitrary context-free grammars. It is likely that a 
PTIME algorithm can be obtained for all context-free 
grammars.

\end{document}